# Nano sized Powder of Jackfruit Seed: Spectroscopic and Anti-microbial Investigative Approach

**T.Theivasanthi [1*], G.Venkadamanickam [2], M.Palanivelu [3], and M.Alagar [4].**

[1,4] Centre for Research and Post Graduate Department of Physics, Ayya Nadar Janaki Ammal College, Sivakasi-626124, Tamilnadu, India.
[2] Rajiv Gandhi Cancer Institute & Research Center, Delhi - 110085, India.
[3] Arulmigu Kalasalingam College of Pharmacy, Krishnankoil - 626126, India.

*Corresponding author. *E-mail*: theivasanthi@pacrpoly.org

**Key Words:** Jackfruit seed powder, FT-IR, EDAX, Anti-bacterial, Specific Surface Area.

***Abstract:***
*This work reports aspect related to nano-sized particles of jackfruit seed. FTIR spectrum was recorded for functional groups analysis and EDAX analysis was done to identify the various elements of the sample. Both FTIR and EDAX analysis results indicated the presence of Starch. FTIR analysis confirmed the availability of anti-microbial Sulphur derivatives compounds. Microbiology assay found that jackfruit seed nanoparticles were effective against Escherichia coli and Bacillus megaterium bacteria. This work also investigated about the dual-function of the sample i.e. food ingredients possessing antimicrobial activities. Specific surface area of bacteria analysis revealed that it played a major role while on reactions with jackfruit seed nanoparticles.*

## 1. Introduction

Natural products are compounds produced by living systems such as plants, animals and microorganisms. They play an important role in the development of drugs. When a natural product is found to be active, it is chemically modified to improve its properties by advances made in synthesis, separation method and biochemical techniques. Naturally existing compounds may be divided into three categories. Firstly, primary metabolites compounds like nucleic acids, amino acids and sugars which occur in all cells and play a central role in the metabolism and reproduction. Secondly, the high-molecular-weight polymeric likes cellulose, lignin and protein which form the cellular structures. Finally, secondary metabolites compounds those are characteristic of a limited range of species. Mostly, drugs are obtained from pure and best behaviour derivation of secondary metabolite natural products. [1].

The *Artocarpus heterophyllus* Lam tree belongs to *Moraceae* family and commonly known as jackfruit. The seeds of this plant have medicinal properties. The oval, oblong or oblong ellipsoid or rounded shape, light brown colour jackfruit seeds are nutritious, rich in potassium, fat, carbohydrates and minerals. Manganese and magnesium elements have also been detected in seed powder [2]. Seeds contain two lectins namely jacalin and artocarpin. Jacalin has been proved to be useful for the evaluation of the immune status of patients infected with human immunodeficiency virus 1 [3].

Nanomaterials are the leading requirement of the rapidly developing field of nanomedicine, bionanotechnology. Nanoparticles usually have better or different qualities than the bulk material of the same element and have immense surface area relative to volume. For centuries, People have used Jackfruit as a food material. Minuscule amounts of jackfruit seed nanoparticles can lend antimicrobial effects to hundreds of square meters of its host material. Jackfruit seeds may therefore be developed into therapeutic agents capable of treating infectious diseases and preventing food contamination by food-borne pathogens. More importantly, the seeds could be processed into dual-functional food ingredients possessing antimicrobial activities.

A survey of literature indicates that not much work has been done on the jackfruit seed. It also appears from the limited published data that no complete FTIR and EDAX investigations on the nano sized particles of jackfruit seed has so far been carried out. For the first time, FTIR and EDAX investigations of the nano sized particles of jackfruit seed are recorded. This work is expected to throw some light on and help further research.

## 2. Experimental Details

We synthesized jackfruit seed nanopowder for these spectroscopic and antibacterial investigations in accordance with our (T.Theivasanthi and M.Alagar) earlier literature procedure [4]. The values of the synthesized sample both size 12 nm and specific surface area 625 $m^2$ $g^{-1}$ which were noted from XRD analysis of the same literature.

Ripe jackfruit seeds were collected from the foot hills of Western Ghat hills at Rajapalayam, Tamilnadu, India  The seeds were cleaned and the white arils (seed coats) were peeled off. Seeds were sun dried for 7 days without remove the thin brown spermoderm. For these experimental purposes, 100g dried seeds were put in a mixer-grinder cum blender which having 550 watts, 17000 rpm rotating speed electrical motor. The seeds were grinded and crushed well and uniformly for 15 minutes with utmost precaution to avoid any contamination and made them as nano-sized powder.

Functional groups analysis of the prepared sample of jackfruit seed nanoparticles were analyzed by SHIMADZU FTIR spectrometer. The FTIR spectrum was recorded for the range of 400 $cm^{-1}$ to 4000 $cm^{-1}$. EDAX analysis was done to identify the presence of various elements. The antibacterial activities of jackfruit seed nanoparticles were studied against *Escherichia coli* and *Bacillus megaterium* by Agar disc diffusion method. Zone of Inhibition (ZOI) was measured and evaluated from this microbiology assay.

## 3. Results and Discussions

### 3.1. FT-IR Analysis

The FT-IR spectrum of Jackfruit seeds are shown in Figure 1. The absorption bands and the wave number ($cm^{-1}$) of dominant peaks obtained from absorption spectrum are presented in Table 1. The observed bands for amines, amides, amino acids indicate the presence of protein and the details are presented in Table 2. It is essential to mention here, Jacalin a major protein is obtained from the jackfruit seeds (tetrameric two-chain lectin combining a heavy chain of 133 amino acid residues with a light β chain of 20-21 amino acid residues). Some other absorption bands indicate the presence of bio-molecules like carbohydrates, polysaccharides and lipids [5].

**Table 1.** FTIR Functional groups analyses

| Vibrational Assignment / Functional Groups | Wave Number ($cm^{-1}$) | Visible Intensity |
|---|---|---|
| N-H $\gamma_{as}$ | 3417.98 | Broad |
| $CH_3$ γ + $CH_2$ $\gamma_{as}$ +$CH_2$ $\gamma_s$ + CH $\gamma_s$ + Conjugate Chelation | 2931.90 | W |
| $CH_2$ γ +N-H γ + O-H γ (- $SO_2$ H) | 2723.58 | VW |
| $NH_3^+$ γ | 2310.80 | VW |
| C=C γ + C=O γ + NH α + $\delta_{as}$ $NH_3^+$ + $NO_2$ $\gamma_{as}$ + N-O γ | 1626.05 | S |
| C=C γ + C=O γ + $COO^-$ $\gamma_{as}$ + N-H α + $\delta_{as}$ $NH_3^+$ + $NO_2$ $\gamma_{as}$ | 1600.97 | S |
| $CH_3$ $\delta_s$ + C-O γ + OH α + C-CHO (skeletal) + $COO^-$ $\gamma_s$ + $NO_2$ $\gamma_s$ + N=O γ + C=S γ + C=F γ | 1383.01 | M |
| $CH_3$ δ(vibrations) + $CH_2$ δ (vibrations) + O-H α + C-O γ + C-CHO (skeletal) + $COO^-$ $\gamma_s$ + C-N γ + $NO_2$ $\gamma_s$ + $SO_2$ $\gamma_{as}$ + C=S γ | 1350.22 | M |
| $CH_2$ β + C-H α + C-O-C $\gamma_{as}$ + C-CO-C (skeletal) + C-CHO (skeletal) + $NH_3^+$ α + N=O γ + $N_3$ $\gamma_s$ + $SO_3$ $\gamma_{as}$ | 1246.06 | VW |
| C-H α + C-O-C $\gamma_s$ + C-CO-C (skeletal) + C-CHO (skeletal) + C-O γ + $NH_3^+$ α + $N_3$ $\gamma_s$ + $SO_2$ $\gamma_s$ + C-F γ | 1155.40 | M |
| C-H α + C-O-C $\gamma_s$ + C-CO-C (skeletal) + C-CHO (skeletal) + C-N γ + N-N γ + S=O γ + C-F γ | 1082.10 | M |
| Skeletal vibrations + Ring γ + C-H α + C-CHO(skeletal) + N-N γ + $SO_3$ $\gamma_s$ + S=O γ + C-F γ | 1018.45 | M |
| Ring γ + CH β + O-H β + N-N γ + N-O γ | 929.72 | VW |
| CH β + C-O-C γ + N-H ω + N-H τ + C-N γ + O-N γ + S=O γ | 858.35 | VW |
| C-Cl γ + C-S γ + O-N γ + N-H ω + N-H τ + OCN(deformation) + C-O-C γ + C-H β | 763.84 | W |
| OCN(deformation) + O-N=O δ + C-Br γ + C-I γ | 576.74 | W |
| S-S γ + C-Br γ + C-I γ | 528.51 | W |
| Abbreviations: α -in-plane bending; β -out-of plane bending; γ -stretching; δ – bending; $\gamma_s$ - symmetric stretching; $\gamma_{as}$ - asymmetric stretching; $\delta_s$ -symmetric bending; $\delta_{as}$ - asymmetric bending; | | |
| M – Medium; S – Strong; W – Weak; VS - Very strong; VW - Very Weak; | | |

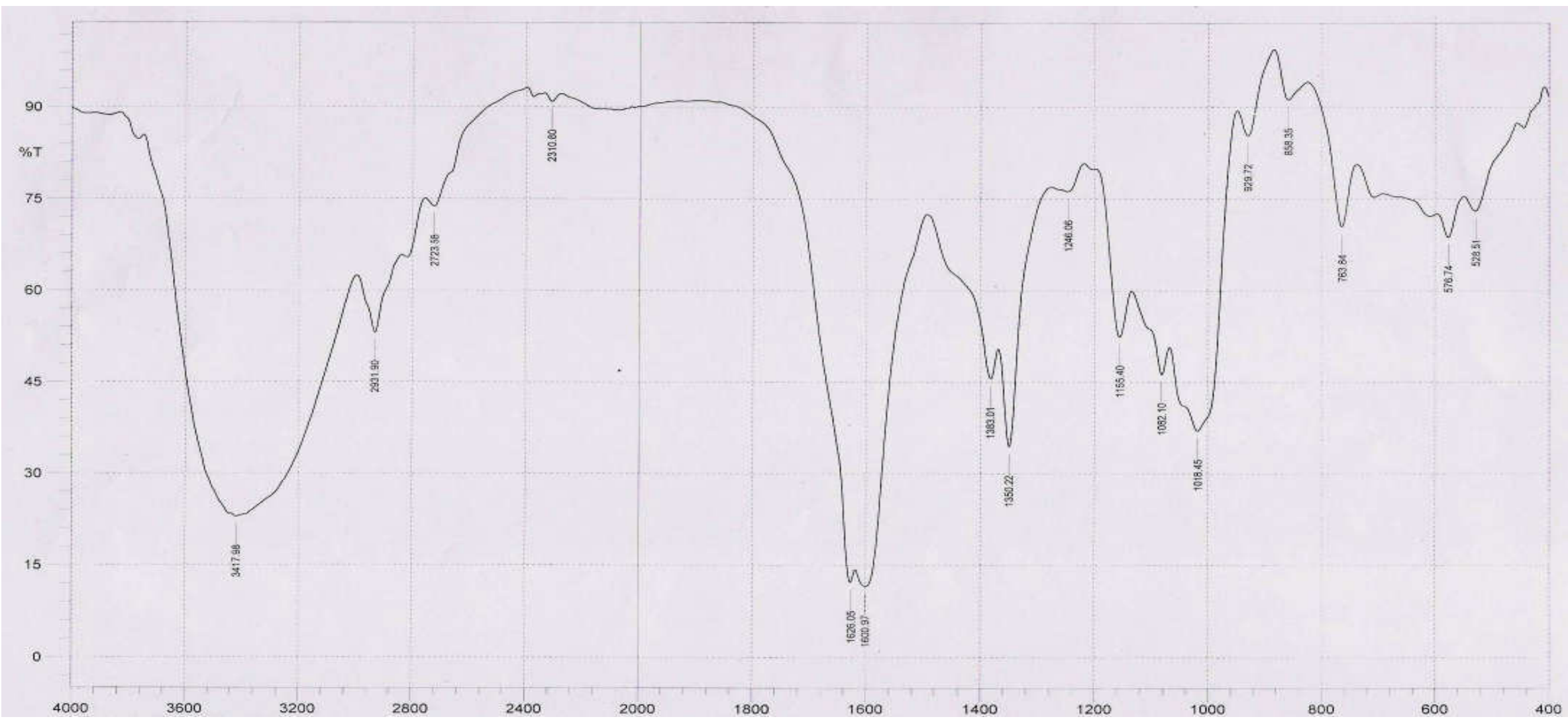

**Figure 1.** Wave number (cm$^{-1}$) of dominant peaks obtained from FT-IR absorption spectrum.

**Table 2.** FTIR Bio – Chemical Compounds analyses

| Bio – Chemical Compounds | | Wave Number (cm$^{-1}$) |
|---|---|---|
| **Amines** | N–H stretching | 3418, 2724, 2311 |
| | N–H bending | 1626, 1601 |
| | C–N stretching | 1350, 1082, 858 |
| **Amides** | N–H stretching | 3418, 2724, 2311 |
| | C–O stretching | 1626, 1601, 1383, 1350, 1155 |
| **Amino acids** | N–H stretching | 3418, 2724, 2311 |
| | N–H bending | 1626, 1601, 1246, 1155 |
| | C–O stretching | 1383 , 1350 |
| **Carboxylic Acids** | O–H stretching | 2724 |
| | C- O stretching | 1626, 1601, 1383, 1350, 1155 |
| **Carbohydrates** | N–H wagging | 858, 764 |
| **Polysaccharides** | C–O–C stretching | 1246, 1155, 1082, 858, 764 |
| **Lipids / Alkanes** | C–H stretching | 2932 |
| **Alkenes** | C=C stretching | 1626, 1601 |
| | C-H out-of plane bending | 930, 858, 764 |
| **Aromatics** | C-H out-of plane bending | 930, 858, 764 |
| **Alcohols, Ethers, Esters, Anhydrides** | C-O stretching | 1155 |
| **Nitrates** | N–H bending | 1626, 1601, 1246, 1155 |
| **Nitro** | N=O | 1350 |
| **Sulfonyl and Sulfonate** | S=O stretching | 1082, 1018, 858 |
| **Chlorate** | C–H stretching | 2932, 2724 |
| **Chloride** | C-Cl | 764 |
| **Fluoride** | C-F | 1383, 1155, 1082, 1018 |
| **Bromide, Iodide** | C-Br, C-I | 577, 529 |

FT-IR report is in agreement with XRD report of our earlier literature (T.Theivasanthi and M.Alagar) and indicating the presence of carbohydrate. Presence of alkanes, alkenes, aromatics, alcohols, ethers, nitrates, sulfonates and organic halogen compounds are also observed. Aromatic compounds indicate existing of flavanoids. Sulphur derivatives compounds are present in jackfruit seeds which exhibit some anti-microbial properties.

The peaks in the finger print region of FTIR spectrum of soluble starch 1383, 1158, 1081, 1021, 931, 860, 575 and 526 are very close with the FTIR spectrum of Jackfruit seed powder [6]. Likewise, FTIR peaks of pure starch 1242.69, 1158.57, 1081.50, 928.96, 860.80, 764.31, 575.28 and 527.83 are also well agreement with the FTIR spectrum of Jackfruit seed powder [7]. The FTIR spectrum of Jackfruit seed powder is matching with the FTIR spectrum of said starch varieties. From this analysis, it is observed that starch is mainly occupying Jackfruit seed powder and this result is also agreeing with the EDAX analysis.

### 3.2. EDAX Analysis

The energy value of each peak is matched with X-ray emission wavelength for non-diffractive analysis and the elements presented in Table 3. Trace elements are estimated by determining the percentage abundance (%) of elements C, O, Mg, Al, P, K, Ti, Fe, Ni and Mo in the sample collected. The EDAX analysis results of such elements are shown in Figure 2. XRD pattern of our earlier literature (T.Theivasanthi and M.Alagar) is mainly showing the properties of starch (a carbohydrate consists only of carbon, hydrogen and oxygen) because of much starch in Jackfruit seeds powder. EDAX analysis is also in well agreement with XRD report and indicating that 94% of carbon and oxygen. Indeed quite a large number of elements are essential to plant and animal (including human) life. If any of these elements is eliminated from our nutrition we would be suffering from one or the other disease / health problem.

**Table 3.** EDAX Quantitaive Analyses - Percentage of Elements

| Element | Net Counts | Weight % | Atom % |
|---|---|---|---|
| ***C*** | 10089 | 45.00 | 54.08 |
| ***O*** | 19390 | 48.98 | 44.19 |
| ***Mg*** | 445 | 0.16 | 0.10 |
| ***Al*** | 667 | 0.20 | 0.11 |
| ***P*** | 584 | 0.16 | 0.08 |
| ***K*** | 1947 | 0.62 | 0.23 |
| ***Ti*** | 998 | 0.49 | 0.15 |
| ***Fe*** | 1445 | 1.07 | 0.28 |
| ***Ni*** | 2988 | 3.09 | 0.76 |
| ***Mo*** | 574 | 0.22 | 0.03 |
| ***Total*** | | 99.99 | 100.00 |

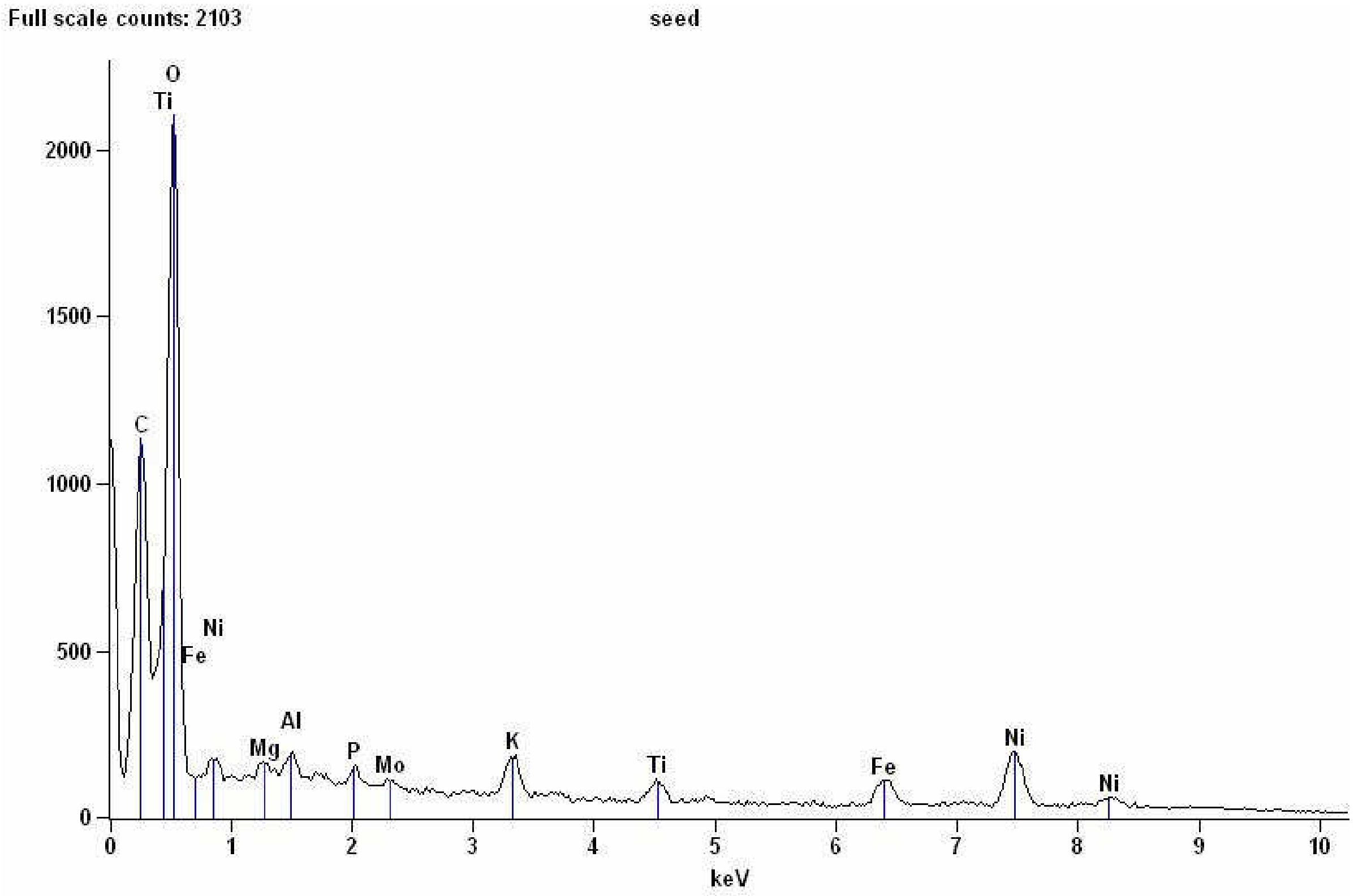

**Figure 2.** EDAX Spectrum showing Elements of Jackfruit seed Nanoparticles.

Carbon is a critical element to all life. It is the second most common element in the human body. It is a constituent of DNA. Oxygen is the most common element in the human body, making up 61 % of the average human's mass. In addition to being a constituent of DNA, it is also plays a role in most other biological compounds. It is a component of water, upon which all life depends. Magnesium is critical to all species, especially plants. Chlorophyll, the pigment that is responsible for photosynthesis, has a single magnesium atom at its center. It is also involved in a number of physiological and biochemical functions. Aluminium is now thought to be involved in the action of a small number of enzymes. Phosphorus is the sixth most common element in the human body. It is a constituent of DNA as well as other biological molecules. Phosphorus is also an important constituent of the bones and teeth.

The potassium ion is used extensively in intercellular fluids. Potassium plays an important role in the growth of plants. Titanium has no known biological use in humans, although it is known to act as a stimulant. In some plants, titanium is used in chemical energy production. Iron is a critical element for life. An iron atom is the center of every hemoglobin molecule and is important for oxygen transport. Nickel is a key metal in several plant and animal enzymes. Molybdenum is an essential trace element to all species. It is especially important to plants to fix nitrogen. Soils that have no molybdenum are generally unfit to support plant life. Molybdenum is important for our cells growth and helpful to control asthma disease.

### 3.3. Anti-Bacterial studies of Jackfruit Seed nanoparticles

Antibacterial activities of jackfruit seed nanoparticles were evaluated by Agar disc diffusion method using Mueller hinton agar medium. A concentration of 50 mcg/ml sample solution was used. Zone of Inhibition (ZOI) was measured from this microbiology assay. The sample showed diameter of zone of inhibition against *E.Coli* 9 mm and *B.megaterium* 7 mm respectively (Figures 3 and 4). In order to disclose the effective factors on their antibacterial activity, many studies have already been focused by various researchers. The crude methanolic extracts of jack seeds exhibited a broad spectrum antibacterial activity [8, 9]. Extract from fresh seeds cures diarrhoea (Bacterial disease) and dysentery [3].

In a solid material, the surface-area-to-volume ratio (SA: V) or Specific Surface Area (SSA) is an important factor for the reactivity that is, the rate at which the chemical reaction will proceed. Materials with large SA: V (very small diameter) reacts at much faster rates than monolithic materials, because more surfaces are available to react. The SSA of cells has an enormous impact on their biology. SSA places a maximum limit on the size of a cell. An increased SSA also means increased exposure to the environment.

***Antibacterial activities evaluation of Jackfruit seed nanoparticles (sample no.3)***

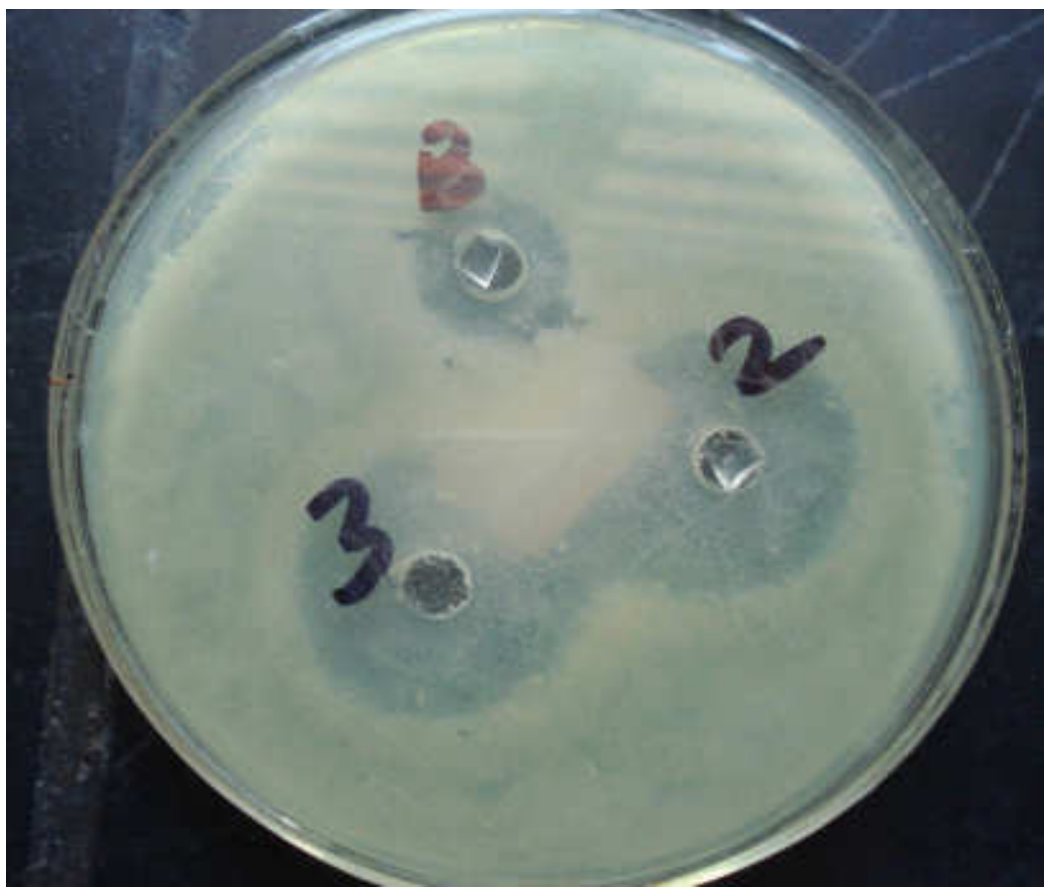

Fig.3. Zone of inhibition diameter against *Escherichia coli* bacteria 9 mm

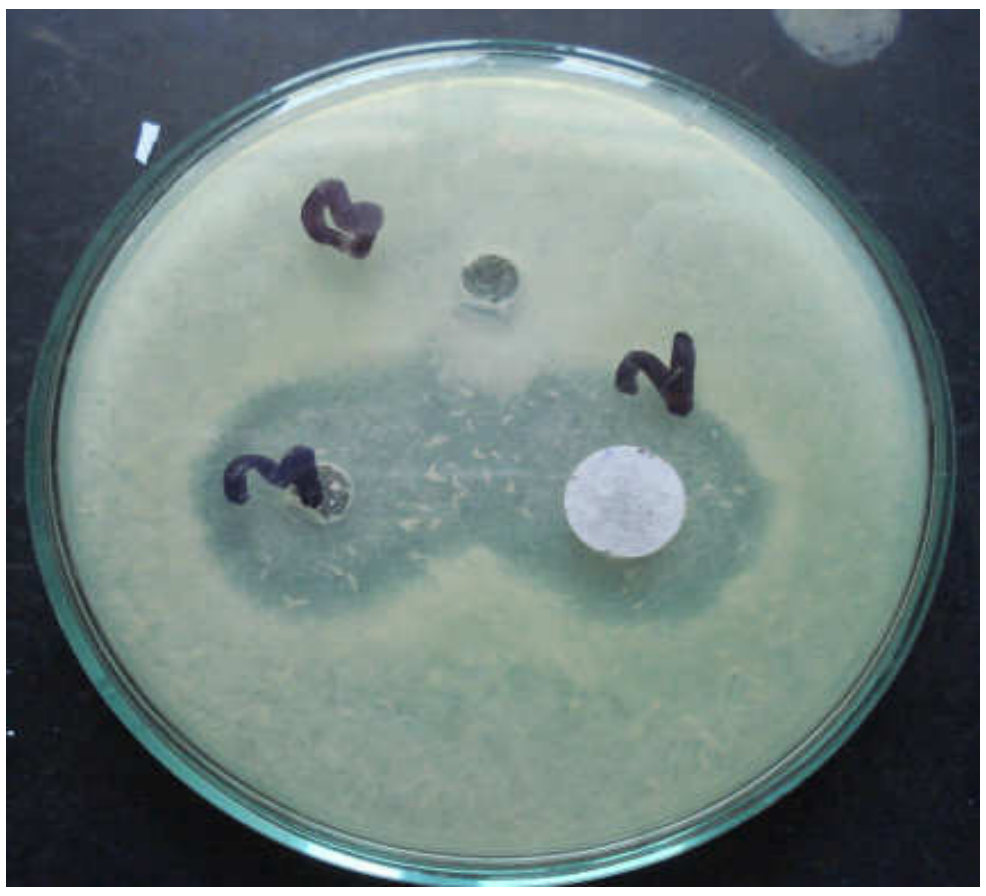

Fig.4. Zone of inhibition diameter against *Bacillus megaterium* 7mm

Greater SSA allows more of the surrounding water to be shifted for nutrients. Increased SSA can also lead to biological problems. More contact with the environment through the surface of a cell increases loss of water and dissolved substances. High SSA also present problems of temperature control in unfavorable environments.

SSA affects the rate at which particles can enter and exit the cell whereas the volume affects the rate at which material are made or used within the cell. These substances must diffuse between the organism and the surroundings. The rate at which a substance can diffuse is given by Fick's law.

$$\text{Rate of Diffusion} \propto \frac{\text{surface area} \times \text{concentration difference}}{\text{distance}} \quad \text{..... (1)}$$

So rate of exchange of substances depends on the organism's surface area that's in contact with the surroundings. Requirements for materials depend on the volume of the organism, so the ability to meet the requirements depends on the SSA.

The Specific Surface Area (SSA) of Jackfruit seed nanoparticles 625 $m^2 g^{-1}$ has been noted from our earlier literature (T.Theivasanthi and M.Alagar). We have made an attempt to study the SSA of bacteria and its reactivity to antibacterial activities of Jackfruit seed nanoparticles. For this study, we have compared SSA of *E.coli* with *B.megaterium* and the details are in Table.4. *E.coli* details (Cell length: 2 µm or $2x10^{-6}$ m, diameter: 0.8 µm or $0.8x10^{-6}$ m, total volume: $1x10^{-18}$ $m^3$, surface area: $6x10^{-12}$ $m^2$, wet weight: $1x10^{-12}$ g, dry weight: $3.0x10^{-13}$ g) has been extracted from The CyberCell Database- CCDB and SSA calculated accordingly [10]. SSA of *B.megaterium* has been noted from the research of Rene Scherrer et al. [11]. From this analysis, we find that *E.coli* has more SSA than *B.megaterium.*

**Table 4.** Specific Surface Area Comparison of Bacteria and Jackfruit Seed Nanoparticles

| Bacteria | | | Jackfruit Seed Nanoparticles | |
|---|---|---|---|---|
| **Name** | **Variety** | **Specific Surface Area** | **Diameter of Zone of Inhibition** | **Specific Surface Area** |
| *Escherichia coli* | Gram (-) | 20.09 $m^2 g^{-1}$ | 9 mm | 625 $m^2 g^{-1}$ |
| *Bacillus megaterium* | Gram (+) | 6.69 $m^2 g^{-1}$ | 7 mm | |

More SSA of *E.coli* increases its exposure to the environment / surroundings in which Jackfruit seed nanoparticles are exist and this condition is unfavourable to *E.coli*. It increases rate of exchange of substances that is in contact with the surroundings of *E.coli*. It leads to more reactions of *E.coli* with Jackfruit seed nanoparticles than *B.megaterium*. Due to its more reactions in unfavourable surroundings results in increased Zone of Inhibition. Bacteria, viruses and fungi all depend on an enzyme to metabolize oxygen to live. Anti-microbial agent such as sulphur derivatives compounds of jackfruit seeds interfere with the effectiveness of the enzyme and inhibit the uptake of oxygen, thereby killing the microbes. This study reveals that the SSA of bacteria plays a major role while reacting with antimicrobial agents.

## 4. Conclusion

Nano-sized particles of jackfruit seed has been prepared and analyzed with FTIR and EDAX techniques successfully. FTIR and EDAX analysis are indicating the presence of Starch (Carbohydrate). FTIR analysis concludes the presence of Sulphur derivatives compounds which contribute anti-microbial activities.

Investigation on the antibacterial effect of nanosized particles of Jackfruit seed against *E.coli* and *B.megaterium* microbes reveals the efficacy of jackfruit seed nanoparticles as an antibacterial agent. SSA of jackfruit seed nanoparticles has been analyzed which concludes that jackfruit seed nanoparticles can lend antimicrobial effects to hundreds of square meters of its host material. Jackfruit seeds may therefore be developed into therapeutic agents capable of treating infectious diseases and preventing food contamination by food-borne pathogens. Jackfruit seeds could be processed into dual-functional food ingredients possessing antimicrobial activities. Likewise, analysis results of SSA of two different bacteria conclude that SSA of bacteria plays a major role while reacting with antimicrobial agents.

This study suggests that jackfruit seed powder has a lot of potential in food, cosmetics, pharmaceuticals, paper, bio-nanotechnology industries, especially its uses as thickener and binding agent. This work throws some light on and helps further research on nano-sized particles of jackfruit seed.

## Acknowledgements

The authors express immense thanks to CECRI, Karaikudi, India for providing EDAX analysis facility, *S.Sivadevi*, *The SFR College for Women*, Sivakasi, India, staff & management of *PACR Polytechnic College*, Rajapalayam, India, *Rajiv Gandhi Cancer Institute & Research Center*, Delhi, India, *Ayya Nadar Janaki Ammal College*, Sivakasi, India and *Arulmigu Kalasalingam College of Pharmacy*, Krishnankoil, India for their valuable suggestions, assistances and encouragements during this work.